\documentclass{PoS}
\PoS{PoS(WC2004)027}
\title{Variables Separation in Gravity}
\ShortTitle{Variables Separation in Gravity}
\author{\speaker{Valery V. Obukhov}
\thanks{The author thanks RFBR  for financial support (grant N~03-01-00105).}
\\
Tomsk State Pedag. Univ., Tomsk, Russia\\
E-mail: \email{obukhov@tspu.edu.ru}}

\author{Konstantin E. Osetrin\\
Tomsk State Pedag. Univ., Tomsk, Russia\\
E-mail: \email{osetrin@tspu.edu.ru}}

\abstract{To solve the problem of exact integration of the field
equations or equations of motion of matter in curved spacetimes
one can use a class of Riemannian metrics for which the simplest
equations of motion can be integrated by the complete separation
of variables method. Here, we consider the particular case of the
class of St\"{a}ckel metrics. These metrics admit integration of
the Hamilton--Jacobi equation for test particle by the complete
separation of variables method. It appears that the other
important equations of motion (Klein--Gordon--Fock, Dirac, Weyl)
in curved spacetimes can be integrated by complete separation of
variables method only for the metrics, belonging to the class of
St\"{a}ckel spaces.}

\FullConference{Fourth International Winter Conference on Mathematical Methods in Physics\\
09 - 13 August 2004\\
Centro Brasileiro de Pesquisas Fisicas (CBPF/MCT), Rio de Janeiro,
Brazil}

\newtheorem{theo}{Theorem}
\newtheorem{defi}{Definition}
\def\pod#1#2{\mathop{#1}_{#2}}
\def\zvez#1#2{\mathop{#1}^{#2}{}}
\usepackage{latexsym}

\begin{document}

\renewcommand{\b}[2]{\beta_#1^{\ #2}}
\newcommand{\g}[2]{\gamma_#1^{\ #2}}
\renewcommand{\a}[1]{\alpha_#1}

\section{Introduction}

One of the main problems of mathematical physics for the gravity
theory is the problem of exact integration of the field equations
or equations of motion of matter. To solve this problem one can
use the class of Riemannian metrics for which the simplest
equations of motion can be integrated by the complete separation
of variables method. Apparently in this class the St\"{a}ckel
metrics are of the same interest. Recall that metric is called the
St\"{a}ckel one if the Hamilton--Jacobi equation
$$
g^{ij}S_{,i}S_{,j}=m^2 \qquad i,j,k,l=1,...n \eqno (1.1)
$$
can be integrated by the complete separation of variables method.
In this case the privileged co-ordinate set $\{u^i\}$ exists for
which complete integral of eq.~(1.1) can be shown in the form
$$
S=\sum_{i=1}^n \phi_i(u^i,\lambda)     \eqno (1.2)
$$
where $\lambda_i$ -- is the essential parameter.

It appears that the other important equations of motion
(Klein--Gordon--Fock, Dirac, Weyl) can be integrated by complete
separation of variables method only for the metrics, belonging to
the class of St\"{a}ckel spaces. That is why the research of this
class of spaces belongs to the one of the important problems of
the mathematical physics. In the present report we consider the
following parts of this problem:
\begin{itemize}
\item The problem of complete separation of variables for the
Hamilton--Jacobi equation.
\item Integration of Einstein equations for the St\"{a}ckel
spaces. \item Conformally St\"{a}ckel spaces.
\item Homogeneous St\"ackel spacetimes.
\end{itemize}


\section{St\"{a}ckel spaces}

The theory of St\"{a}ckel spaces has been developed by many
authors.
Let us recall the main statements
and enumerate the main theorems of the St\"{a}ckel spaces theory.
\begin{defi}
Let $V_n$ be a $n$--dimensional Riemannian space with metric
tensor $g_{ij}$. The Hamilton -- Jacobi equation can be integrated
by complete separation of variables method if co-ordinate set
$\{u^i\}$ exists for which complete integral can be presented in
the form (1.2).
\end{defi}
\begin{defi}
$V_n$ is called the St\"{a}ckel space if the Hamilton--Jacobi
equation (1.1) can be integrated by complete separation of
variables method.
\end{defi}
The following theorem was proved by V.N.~Shapovalov
\cite{4.}--\cite{5.}.
\begin{theo}
Let $V_n$ be the St\"{a}ckel space. Then $g_{ij}$ in privileged
co-ordinate set can be shown in the form
$$
g^{ij}=(\Phi^{-1})^\nu_n G^{ij}_\nu,
\qquad
G^{ij}_\nu=G^{ij}_\nu(u^\nu), \qquad
\Phi^\nu_\mu=\Phi^\nu_\mu(u^\mu) \eqno (2.1)
$$
$$
G^{ij}_\nu=\delta^i_\nu\delta^j_\nu\varepsilon_\nu(u^\nu) +
(\delta^i_\nu\delta^j_p+\delta^j_\nu\delta^i_p)G^{\nu p}_\nu +
\delta^i_p\delta^j_qG^{pq}_\nu,
\qquad
p,q=1,...N, \quad \nu,\mu=N+1,...,n.
$$
where  $\Phi^\nu_\mu(u^\mu)$ -- is called the St\"{a}ckel matrix.
\end{theo}
It is assumed that summation is performed over repeated
superscripts and subscripts provided the symbol $ns(i,j,...)$ (no
summation over the indices given in the brackets) does not occur
on the right of the formula. One can show that the geodesic
equations of St\"{a}ckel spaces admit the first integral that
commutes pairwise with respect to the Poisson bracket
$$
\pod X\mu=(\Phi^{-1})^\nu_\mu H_\nu, \quad H_\nu=\varepsilon_\nu
p^2_\nu + 2G^{\nu p}_\nu p_pp_\nu + h^{pq}_\nu p_pp_q,
\qquad
\pod Yp=\pod Yp{}^ip_i.  \eqno(2.2) $$ Thus for a covariant
characterization of St\"{a}ckel spaces it is sufficient to find
determining properties of the integrals (2.1) in an arbitrary
co-ordinate system $\{x^i\}_n$. Let us write the functions $\pod
X\nu$, $\pod Yp$ in the form
$$
\pod X\nu=\pod X\nu {}^{ij}p_ip_j,\qquad \pod Yp=\pod Yp{}^ip_i,
\qquad \mbox{then} \qquad
\pod X\nu{}_{(ij;k)}=\pod Yp{}_{(i;j)}=0
\eqno (2.3)
$$
(the semicolon denotes the covariant derivative and the brackets
denote symmetrization). Therefore $\pod Yp{}^i$, $\pod
X\nu{}^{ij}$ are the components of vector and tensor Killing
fields respectively.


\begin{defi}
Pairwais commuting vector $\pod Yp{}^i$, where $p=1,...N$ and
tensor $\pod X\nu{}^{ij}$, where $\nu=N+1,...n$ Killing fields
form a complete set of the type $(N.N_0)$ if
$$
B^{pq}\pod Yp{}^i\pod Yq{}^j + B^\nu\pod X\nu{}^{ij}=0 \quad
\Longrightarrow \quad B^{pq}=B^\nu=0 \eqno (2.4)
$$
$$
rank||\pod Yp{}^i\pod Yq{}_i||=N-N_0   \eqno(2.5)
$$
$$
\pod X\nu{}^{ik}\pod X\mu{}^j{}_k=C^{pq}_{\nu\mu} \pod Yp{}^i\pod
Yp{}^j + C^\tau_{\nu\mu}\pod X\tau{}^{ij}, \eqno (2.6)
$$
$$
C^\tau_{\nu\mu}={\Phi^\tau_\rho(\Phi^{-1})^\rho_\nu(\Phi^{-1})^\rho_\mu
/ (\Phi^{-1})^\rho_n}
$$
$$
\pod X\nu{}^{ij}\pod Yp{}_j=C^q_{\nu p}\pod Yq{}^i. \eqno (2.7)
$$
\end{defi}
\begin{theo}
A necessary and sufficient geometrical criterion of a St\"{a}ckel
space is the presence of a complete set of the type ($N.N_0$).
\end{theo}
In other words the Hamilton - Jacobi equation can be integrated by
the complete separation of variables method if and only if the
complete set of the first integrals exists.

\section{ St\"{a}ckel spaces and field equations of the theories of gravity}

The metrics of the St\"{a}ckel spaces can be used for integrating
the field equations of General Relativity and other theories of
Gravity. Note that such famous solutions as Schwarcshild, Kerr,
NUT, Friedman and others belong to the class of St\"{a}ckel
spaces. Apparently the first papers devoted to the problem of
classification of the St\"{a}ckel spaces satisfying the Einstein
equations were published by B.Carter \cite{11.}. Later in our
paper \cite{14.} it has been found the complete classification of
the special St\"{a}ckel electrovac spaces. In other words all
St\"{a}ckel spaces satisfying the Einstein--Maxwell equations for
the case when potentials $A_i$ admit complete separation of
variables for Hamilton--Jacobi equation (2.8) have been found. In
our paper \cite{28.} the classification problem has been solved
for the case when $A_i$ are arbitrary functions and spaces are
null (types (N.1)). In our paper \cite{27.} all electrovac
spacetimes admitting diagonalization and complete separation of
variables for the Dirac--Fock--Ivanenko equation were found.

One of the complicated problems of the modern mathematical physics
is the integration problem of the Einstein--Dirac equations. Using
the Newman--Penrose formalism one can present these equations in
the form
$$
R_{ij}-{1\over2}g_{ij}R=4\pi GT_{ij},
$$
$$
\nabla_{ab'}\xi^a=m_0\eta_{b'},\qquad
\nabla_{ab'}\eta^a=m_0\xi_{b'}   \eqno (3.1)
$$
where
$$
T_{ij}=Z^a_iZ^b_j\sigma^{AB'}_a\sigma^{CD'}_bT_{AB'CD'},
\qquad
 Z^a_i=(l_i,n_i,m_i,\bar m_i)
$$
$$
T_{AB'CD'}=ik(\xi_{D'}\nabla_{AB'}\xi_C + \xi_B\nabla_{CD'}\xi_A
-
\xi_C\nabla_{AB'}\xi_{D'} - \xi_A\nabla_{CD'}\xi_{B'}-
\eta_{D'}\nabla_{AB'}\eta_C - \mbox{} $$ $$
\eta_B\nabla_{CD'}\eta_A + \eta_C\nabla_{AB'}\eta_{D'} +
\eta_A\nabla_{CD'}\eta_{B'})
$$
and $\nabla_{AB}$ - spinor derivative, $\sigma^{AB'}_a$ -
Infeld-Wan der Varden symbols.

Using spaces for which equation (3.1) can be integrated by the
complete separation of variables and separated solutions of the
Dirac equation one can transform Einstein--Dirac equations to the
set of functional equations. The first papers devoted to the
classification problem for the Einstein--Dirac equations were done
by Bagrov, Obukhov, Sakhapov \cite{29.}. The St\"{a}ckel spaces of
type (3.1) for Einstein--Dirac and Einstein--Weyl equations have
been studied. Appropriate solutions have been obtained. They
contain arbitrary functions depending on null variable only.

The problem of classification of St\"{a}ckel spaces for other
theories of gravity for the first time was considered in papers
\cite{30.}--\cite{32.}. The following theories have been
considered

\begin{itemize}

\item[1.] Brans--Dicke theory. The field equations have the form
$$
R_{ij}-{1\over2}g_{ij}R=
{8\pi\over\phi}T_{ij}-{\omega\over\phi^2}(\phi_{;i}\phi_{;j}-
{1\over2}g_{ij}\phi_{;k}\phi^{;k})-
{1\over\phi}(\phi_{;ij}-g_{ij}\Box\phi)   \eqno (3.2)
$$
$$
\Box\phi={8\pi\over3+2\omega}T^i{}_i, \quad
\Box=g^{ij}\nabla_i\nabla_j, \quad \omega=const.
$$
The case of Einstein--Maxwell equations and the metric has type
(N.1) has been considered in paper \cite{30.}. All appropriate
solutions have been obtained.

\item[2.] The problem for the multiscalar--tensor theory for
which field equations have the form 
$$
\zvez R*_{\nu\mu}=2<\phi_{,\nu}\phi_{,\mu}>+8\pi G(\zvez
T*_{\nu\mu}- {1\over 2}\zvez g*_{\nu\mu}\zvez T*),  
\quad
\zvez\Box *\phi^A+\gamma^A{}_{BC}\phi^A_{,\nu}\phi^B_{,\mu} \zvez
g*^{\nu\mu}=4\pi G\gamma^{AB}
{\partial\alpha\over\partial\phi^B}\zvez T* \eqno (3.3)
$$
was considered in paper \cite{32.}. Here $\zvez g*_{\nu\mu}$,
$\zvez R*_{\nu\mu}$ are metric tensor and Ricci tensor of the
space which conformal to space-time,
$$
\zvez g*_{\nu\mu}=\alpha^2g_{\nu\mu},\qquad \alpha=\alpha(\phi^A),
\qquad
<\phi_\nu\phi_\mu>=\gamma_{AB}\phi^A_{,\nu}\phi^B_{,\mu},
$$
where $\phi^A$ - are scalar fields,
$\gamma_{AB}=\gamma_{AB}(\phi^C)$ can be considered as a metric
tensor of the $n$ - dimensional space of scalars,
$$
\gamma^C_{AB}={1\over 2}\gamma^{CD}(
{\partial\gamma_{AD}\over\partial\phi^B} +
{\partial\gamma_{DB}\over\partial\phi^A} -
{\partial\gamma_{AB}\over\partial\phi^C}),
\qquad
\zvez\Box *\phi^A\equiv (\sqrt{|\zvez g*|}\zvez
g*^{\nu\mu}\phi^A_{,\nu})_{,\mu} /\sqrt{|\zvez g*|}
$$

\item[3.] The classification problem for the Einstein--Vaidya
equations. Let the stress--energy tensor have the form
$$
T_{ij}=T^{(e)}_{ij} + a(x)l_il_j,\qquad l_il^i=0 \eqno (3.4)
$$
then Einstein equations can by written in the following way
$$
R_{ij}-{1\over2}g_{ij}R=4\pi G(T^{(e)}_{ij}+al_il_j). \eqno (3.5)
$$
If $T^{(e)}_{ij}$ has the form for electrovac, and $F_{ij}$
satisfies the Maxwell equations (2.12) the solutions of the
equations (3.5) are electrovac ones. For these equations the
classification problem was solved in paper \cite{28.} for the case
when the complete set has type (N.1) (null case). In other words
all metrics and electromagnetic potentials satisfying equations
(3.5) provided that Hamilton--Jacobi equation (1.1) can be
integrated by the complete separation of variables method for null
St\"{a}ckel spaces have been found.
\end{itemize}


\section{Conformally St\"{a}ckel spaces}

Let us consider the Hamilton--Jacobi equation for a massless
particle
$$
g^{ij}S_{,i}S_{,j}=0 \eqno (4.1)
$$
Obviously this equation admits complete separation of variables
for a St\"{a}ckel space. Yet one can verify that if $g^{ij}$ has
the form
$$
       g_{ij} = \tilde g_{ij} (x) \exp{2\omega(x)} \eqno (4.2)
$$
where $\tilde g_{ij}$ is a metric tensor of the St\"{a}ckel space,
then eq.(4.1) can be solved by complete separation of variables
method too. In paper \cite{5.} it was proved that (4.2) is
necessary and additional condition of the complete separation of
variables. Note that conformally St\"{a}ckel spaces play important
role when massless quantum equations are considered (f.e.
conformal invariant Chernikov--Penrose equation, Weyl's equation
etc.). That is why the problem of investigation of Einstein spaces
which admits complete separation of variables in eq.(4.1) is
exceptionally interesting. Apparently the first attempt to
consider this problem was taken in paper \cite{34.}. The next step
was done in paper \cite{35.}, where some of metrics belonging to
conformal St\"{a}ckel spaces of type (N.1) was studied. The
problem of classification of conformally St\"{a}ckel spaces
satisfying the Einstein equation
$$
R_{ij}=\Lambda g_{ij}, \qquad \Lambda=const \eqno(4.3)
$$
is more difficult than appropriate problem for the St\"{a}ckel
spaces. To obtain the functional equations from eq. (4.3) one has
to use the integrability conditions. These conditions was found by
Brinkman \cite{33.}.
Let us denote $V_n$ the Riemannian space with metric tensor
$g_{\alpha\beta}$, $\tilde V_n$ be an Einstein's space with metric
tensor $\tilde g_{\alpha\beta}$. $\tilde R_{\alpha\beta}, \tilde
R_{\alpha\beta\gamma\delta}, \tilde R$  are components of Ricci
tensor , Riemann tensor and scalar curvature respectively for the
space $\tilde V_n$, and $R_{\alpha\beta},
R_{\alpha\beta\gamma\delta}, R$ are those tensors for the space
$V_n$. Moreover we denote:
$$  T _{\alpha\beta}={1\over{n-2}}\left( R_{\alpha\beta}
-{{R g_{\alpha\beta}} \over{2(n-1)}} \right) ,
\qquad
W =
{1\over2}(\nabla\omega)^2 - {\Lambda\over{2(n-1)}} \exp{2\omega}
$$
where
$\omega_{;\alpha}\equiv\nabla_{\alpha}\omega$ are the covariant
derivatives in $V_n$.

We can write (4.3) in the form
$$ \omega_{,\alpha ;\beta} - \omega_{,\alpha} \omega_{,\beta} +
 W  g_{\alpha\beta} + T _{\alpha\beta}= 0.   \eqno (4.4)
$$

Brinkman has shown that integrability conditions of eq.~(4.4) have
the form
$$ \omega_{,\delta}C^\delta{ }_{\alpha\beta\gamma} =
S_{\alpha\beta\gamma},          
\qquad\mbox{where} \quad S_{\alpha\beta\gamma} \equiv
T_{\alpha\gamma ;\beta } - T _{\alpha\beta ;\gamma} \eqno (4.5)
$$
where $C_{\alpha\beta\gamma\delta}$ are the components of Weyl
tensor.
In our paper \cite{35.} eqs.(4.5) have been presented in a more
simple form. To simplify them we use the consequence from Bianchi
identities
$$ R^{\sigma}{ }_{\alpha\beta\gamma ;\sigma} =
   R_{\alpha\beta ;\gamma} -  R_{\alpha\gamma ;\beta}.  \eqno (4.6)
$$
One can take out the following correlation
$$ C^{\alpha}{ }_{\beta\gamma\delta ;\alpha} =
R^{\alpha}{ }_{\beta\gamma\delta ;\alpha} + S_{\beta\gamma\delta}
-{1\over2(n-1)}(g_{\beta\gamma} R_{,\delta} -
g_{\beta\delta}R_{,\gamma}).
                                                            \eqno (4.7)
$$
That is why
$$ C^{\delta}{ }_{\alpha\beta\gamma ;\delta} =
   -(n-3)S_{\alpha\beta\gamma}             \eqno (4.8)
$$
and Brinkman's conditions can be presented in the form
$$ \omega_{,\delta}C^{\delta}{ }_{\alpha\beta\gamma} =
   -{1\over (n-3)}C^{\delta}{ }_{\alpha\beta\gamma ;\delta},
$$
finally
$$\nabla_\delta \left (C^{\delta}{ }_{\alpha\beta\gamma}
\exp{(n-3)\omega}\right) = 0.    \eqno (4.9)
$$
Using (4.9) we have proved the following theorem \cite{35.}
\begin{theo}
Let $g_{ij}$ be the metric tensor of the St\"{a}ckel space of type
(N.1). Then Einstein space conformal to $\tilde V_4$ admits the
same Killing vectors as $V_4$.
\end{theo}
Moreover one can prove the following statement.
\begin{theo}
Let $\tilde V_n$ is conformally St\"{a}ckel space of type (N.1)
($N\geq 2$) satisfying the Einstein equation (4.3). Then
Hamilton--Jacobi equation (1.1) admits the complete separation of
variables.
\end{theo}
In other words all null conformally St\"{a}ckel Einstein spaces
belong to the class of the null ones. Nontrivial null conformally
St\"{a}ckel solutions of the Einstein equations belong only to
(1.1)-type.


\section{Homogeneous St\"ackel spacetimes}

One of the problems of cosmology is the problem of finding exact
realistic models. The most interesting for cosmology are
space-homogeneous models, which are known to admit a
3-parametrical transitive group of motions with space-like orbits.

On the
other hand, it is known that geodesic equations in the form of
Hamilton-Jacobi can be integrated by the complete separation of
variables method if they admit a first integrals which are linear
and quadratic with respect to the momenta
$
\mathop Y_p=\mathop Y_p{}^ip_i,
$
$
\mathop X_\nu=\mathop X_\nu {}^{ij}p_ip_j, $
where $Y^i$ -
Killing vectors and $X^{ij}$ - Killing tensors.

Thus there is a problem of finding a subclass of homogeneous
space-times admitting complete sets of integrals of motion. In
other words, a space-time with a complete set must admit a
3-parametrical transitive group of motions with space-like orbits.

There are 7 types of complete sets for space-times with the
signature $(-,+,+,+)$. In our papers \cite{3a}--\cite{4a} we find
a complete classification of homogeneous St\"ackel space-times
(3.1) type.

Let us consider space-times with a complete set of type (2.1).
The contravariant metric tensor of the St\"ackel space of type
(2.1) in a privileged coordinate system can be written as
(signature $(-,+,+,+)$)
\begin{equation}
g^{ij} = \frac 1{\Delta} \left(\matrix{1&0&0&0\cr 0&0&f(x^1)&1\cr
   0&f(x^1)&c(x^0,x^1)&b(x^0)\cr 0&1&b(x^0)&a(x^0)\cr}\right)
\end{equation}
$\mbox{where}\ \ \Delta=d_0(x^0)+d_1(x^1), \quad c =
c_0(x^0)+c_1(x^1)$
\[
\det{g^{ij}}=-\frac{D}{\Delta^4},\qquad D=a\,f^2-2\,b\,f+c>0
\]
The complete set of type (2.1) includes the following Killing
vectors:
\begin{equation}
X_1=(0,0,0,1)\ ,\quad X_2=(0,0,1,0)
\end{equation}
The vector $X_2$ is space-like because $D>0$; however, the
restriction of the metric to the orbits of subgroup $X_1,X_2$ is
degenerated. It is a general property of St\"ackel spaces with
$N_0 \neq 0$, which are called null spaces. The breaking of
space-likeness of the orbits of the complete set subgroup demands
the introduction of an additional Killing vectors -- $X_3$ and
$X_4$.

The commutation relations are
\begin{equation}
\begin{array}{l}
\left[X_1,X_2\right]=0\\
\left[X_1,X_a\right]=\alpha_a\,X_1+\beta_a{}^b\,X_b,\qquad a,b=3,4\\
\left[X_2,X_a\right]=\gamma_a{}^2\,X_2+\gamma_a{}^b\,X_b\\
\left[X_3,X_4\right]=\gamma_5\,X_3+\gamma_6\,X_4+\gamma_7\,X_2
\end{array}
\end{equation}
To find algebraic classes, the vectors of this subgroup can be
subjected to the linear transformation
\begin{equation} \tilde{X}_\alpha=S_\alpha{}^\beta\,X_\beta,\qquad \alpha, \beta=2..4
\label{tra1}\end{equation} Also, one can use a coordinate
transformation keep the form of the metric tensor
\begin{equation} \begin{array}{l}
 \tilde{x^p}=\alpha^p\,x^p \qquad\qquad\quad p, q = 0,1 \\
 \tilde{x^\nu}=\alpha^\nu+\beta_{\mu}{}^\nu(x^p)\,x^\mu\qquad
 \qquad \mu,\nu=2,3\end{array} \label{tra2}\end{equation}
Applying the admissible transformations and the Jacobi identities
to the commutation relations, integrating them and Killing
equations we obtain complete classification of this space-times
(all metrics and Killing vectors) and establish the following
result.
\begin{theo}
Homogeneous St\"{a}ckel space of type (2.1) have only Petrov's
type III or type N metrics with constant scalar curvature. This
spacetimes can not be Bianchi type VIII or IX.
\end{theo}

This work partially supported by RFBR, grant N~03-01-00105.

\end{document}